\documentclass[11pt]{article}
\usepackage{amsmath}
\usepackage{amsfonts}
\usepackage{amssymb}
\usepackage{graphicx}
\usepackage{bm}
\usepackage{color}
\usepackage{amscd}

\def\bea{\begin{eqnarray}}
\def\eea{\end{eqnarray}}

\begin{document}
\begin{center}
\LARGE {\bf BMS type symmetries at null-infinity and near horizon of non-extermal black holes }
\end{center}

\begin{center}
{M. R. Setare \footnote{E-mail: rezakord@ipm.ir}\hspace{1mm} ,
H. Adami \footnote{E-mail: hamed.adami@yahoo.com}\hspace{1.5mm} \\
{\small {\em  Department of Science, University of Kurdistan, Sanandaj, Iran.}}}\\

\end{center}

\begin{center}
{\bf{Abstract}}\\
In this paper we consider a generally covariant theory of gravity, and extend the generalized off-shell ADT current such that it becomes conserved for field dependent (asymptotically) Killing vector field. Then we define the extended off-shell ADT current and the extended off-shell ADT charge. Consequently, we define the conserved charge perturbation by integrating from the extended off-shell ADT charge over a spacelike codimension two surface. Eventually, we use the presented formalism to find the conserved charge perturbation of an asymptotically flat spacetime. The conserved charge perturbation we obtain is exactly matched with the result of the paper \cite{6'}. These charges are as representations of the
$BMS_4$ symmetry algebra. Also, we find that the near horizon conserved charges of a non-extremal black hole with extended symmetries are the Noether charges. For this case our result is also exactly matched with that of the paper \cite{15}.
\end{center}

\section{Introduction}
It is well known that the group of asymptotic symmetries of asymptotically flat space-times
 at future null infinity is the BMS group \cite{1',2',10}. The BMS symmetry algebra in $n$ space-time dimension consists
 of the semi-direct sum of the conformal Killing vectors of a $(n-2)$-dimension sphere acting on the ideal of infinitesimal
 supertranslations \cite{4',5'}. So, in 4-dimensions, the asymptotic symmetry group at null infinity of asymptotically
flat space-times is not the Poincare group. In this case, the symmetry algebra is an extension of the Poincare algebra,
in which translations are replaced by supertranslations, and contains two copies of the Virasoro algebra \cite{6'}.
 In contrast to 3 and 4 dimensions, in higher dimensions the supertranslations reduce to the usual translation.
 Thus the asymptotical symmetry algebra of asymptotically flat space-times at the future null infinity for $n>4$ dimensions is just the Poincare algebra \cite{7',8',5'}. The infinite-dimensional supertranslation subgroup of $BMS_4$ generates arbitrary angle dependent translations of retarded time. Ashtekar has investigated the implications of the supertranslations in the context of asymptotic quantization \cite{9',10'}. Recently, Donnay et al \cite{11'}, have shown that the asymptotic symmetries close to the horizon of the non-extremal black hole solution of the three-dimensional Einstein gravity in the presence of a negative cosmological term, are generated by an extension of supertranslations. They have shown that for a special choice of boundary conditions, the near region to the horizon of a stationary black hole presents a generalization of supertranslation, including a
semidirect sum with superrotations, represented by Virasoro algebra (see also \cite{b}). More recently, we have studied the behaviors and algebras of the symmetries and conserved charges near the horizon of the non-extremal black holes in the context of the so-called Generalized Minimal Massive Gravity \cite{30'}, proposed in Ref. \cite{200'}. In an interesting paper \cite{12'}
 Strominger has studied the $BMS_4$ invariance of gravitational scattering. He has discussed that in
 a finite neighborhood of the Minkowski vacuum, classical gravitational scattering is in fact BMS-invariant.
 He has demonstrated BMS invariance of the S-matrix, and has shown that the supertranslation invariance
 implies energy conservation at every angle. In extension of AdS/CFT correspondence to the flat space holography,
 the BMS algebra has been investigated very much in recent years \cite{13',14',15',16',17',18',19',4',20',11,6',21'}. Since the $BMS_4$
 algebra is an extension of the Poincare algebra, the asymptotically flat space-time in 4-dimensions is dual
 to an extended conformal field theory. The $BMS_4$ charge algebra has been studied in \cite{6'}. The authors
 of \cite{6'} have used the covariant approach in order to obtain surface charges and their algebra
 (see also \cite{22',23',24'}). The BMS$_{4}$ Surface-charge algebra on the null infinity of asymptotically flat spacetime has been investigated via the Hamiltonian formalism in the reference \cite{a'}. The  $BMS_4$ group leads to the conserved charges, a part of these conserved charges
 are associated to the Poincare group, another part is an infinite number of supermomentum charge associated with
 supertranslations. In the papers \cite{20',5',6'} Barnich and Troessaert have discussed the vector fields called
 "superrotations", which correspond to the infinitesimal symmetries but cannot be exponentiated to lead smooth
 finite diffeomorphisms. Recently, Flanagan and Nichols have computed the superrotation charges, and have shown that
 these charges which are called "super center-of-mass" by them are in general finite  \cite{25'}.\\
 In this paper we are going to show that the BMS symmetries appear at null infinity of asymptotically flat four-dimensional space-times.
What we find in this case is exactly matched with the result of the paper \cite{6'}, but here, we obtain this result by a different approach.
   On the other hand, the near horizon geometry of non-extermal black hole solutions of a generally covariant theory of gravity exhibits an infinite-dimensional symmetry which is not exactly $BMS_4$ \cite{15}. In fact, at the horizon we find the so-called $BMS^H_4$ algebra provided in reference \cite{15}.\\
  In this paper at first we consider a generally  covariant theory of gravity in $D$ dimensions. Then we try to find the quasi-local conserved charges corresponding to the field-dependent Killing vector fields. We assume that the diffeomorphism generator $\xi$ depends on the dynamical fields which appear in the metric. Then by using the ADT method \cite{1000,1001,1002}, developed  in \cite{1,2,3} (for the recent works see \cite{29,150,40}), we obtain the extended off-shell ADT current. Afterward, we define the perturbation of the conserved charge by integrating of the extended off-shell ADT charge over a space-like codimension two surface. As an application of this method for computation of conserved charges, when the Killing vector fields are dependent to the dynamical fields in the metric, and in direction of our aim, we consider an asymptotically flat line element where the components of the metric are functions of coordinates. Our result for conserved charge perturbation is exactly matched with that of the paper \cite{6'}. These charges are as representations of the
$BMS_4$ symmetry algebra.  Very recently the authors of \cite{15} have shown that the non-extremal black holes in four-dimensional general relativity
exhibit an infinite-dimensional symmetry in their near horizon region. The algebra they have found contains two
sets of supertranslations currents, besides, it contains two
sets of Virasoro currents which are in semi-direct sum with the supertranslations. Due to presence of  two
sets of supertranslations currents, this algebra is not exactly $BMS_4$ symmetry algebra which includes two copies of Virasoro algebra and one set of supertranslations. In section 3 we try to find the expression of the conserved charges associated to the near horizon symmetry of the non-extermal black hole solution of general relativity in 4 dimensions by the Neother method. Our result for this case is also exactly matched with that of the paper \cite{15}.
\section{Quasi-local conserved charges correspond to the field dependent Killing vector fields}
The action of a generally covariant theory of gravity in $D$ dimensions is given by
\begin{equation}\label{1}
  I= \frac{1}{16 \pi G} \int d^{D} x \sqrt{-g} \mathcal{L},
\end{equation}
where $G$ is the gravitational constant and $\mathcal{L} = \mathcal{L} (g_{\mu \nu} , R, R_{\mu \nu} R^{\mu \nu},\dots)$ is the Lagrangian density of a generally covariant theory of gravity. By varying Eq.\eqref{1} with respect to metric $g_{\mu \nu}$ we have
\begin{equation}\label{2}
\delta \left( \sqrt{-g} \mathcal{L} \right) = \sqrt{-g} \mathcal{E} ^{\mu \nu} \delta g_{\mu \nu} + \partial _{\mu} \Theta ^{\mu} (g ; \delta g),
\end{equation}
where $\mathcal{E} ^{\mu \nu}=0$ are equations of motions and $\Theta ^{\mu} (g ; \delta g)$ is the surface term. The variation of metric under a diffeomorphism generated by the vector field $\xi ^{\mu}$ is $\delta _{\xi} g_{\mu \nu} = \nabla _{\mu} \xi _{\nu} + \nabla _{\nu} \xi _{\mu}$. By supposing that the variation in Eq.\eqref{2} is due to a diffeomorphism generated by the vector field $\xi ^{\mu}$ we find that
\begin{equation}\label{3}
 \delta _{\xi} \left( \sqrt{-g} \mathcal{L} \right) = 2 \sqrt{-g} \mathcal{E} ^{\mu \nu} \nabla _{\mu} \xi _{\nu} + \partial _{\mu} \Theta ^{\mu} (g ; \delta _{\xi} g).
\end{equation}
It is known that $\sqrt{-g}$ is a scalar density of weight $+1$ and the Bianchi identity is given by $\nabla_{\nu} \mathcal{E} ^{\mu \nu} =0$, then Eq.\eqref{3} can be written as
\begin{equation}\label{4}
  \partial_{\mu} J^{\mu} =0,
\end{equation}
where $J^{\mu}= J^{\mu} (g ; \xi)$ is an off-shell conserved current and it is given by
\begin{equation}\label{5}
  J^{\mu}= \Theta ^{\mu} (g ; \delta _{\xi} g) - \sqrt{-g} \mathcal{L} \xi ^{\mu} + 2 \sqrt{-g} \mathcal{E} ^{\mu \nu} \xi _{\nu}.
\end{equation}
By virtue of the Poincare lemma, one can write
\begin{equation}\label{6}
  J^{\mu} (g ; \xi) = \partial _{\nu} K^{\nu \mu} (g ; \xi),
\end{equation}
where $K^{\mu \nu} = -K ^{\nu \mu}$. Now, we assume that the diffeomorphism generator $\xi$ depends on the dynamical fields which are appear in the metric.
By varying Eq.\eqref{5} with respect to the dynamical fields we find that

$$ \partial _{\nu} \left( \hat{\delta} K^{\nu \mu}(g ; \xi) - K^{\nu \mu}(g ; \hat{\delta} \xi)- 2 \xi ^{[\nu} \Theta ^{\mu]}(g ; \hat{\delta} g ) \right)$$
$$ = \hat{\delta} \Theta ^{\mu}(g ; \delta _{\xi} g ) - \delta _{\xi} \Theta ^{\mu}(g ; \hat{\delta} g ) - \Theta ^{\mu}(g ; \delta _{\hat{\delta} \xi} g ) $$
\begin{equation}\label{7}
  + 2 \sqrt{-g} \left( \hat{\delta} \mathcal{E} ^{\mu \nu} \xi_{\nu} + \mathcal{E} ^{\mu \nu} \hat{\delta} g_{\nu \lambda} \xi ^{\lambda} - \frac{1}{2} \xi ^{\mu} \mathcal{E} ^{\alpha \beta} \hat{\delta} g_{\alpha \beta} + \frac{1}{2} g ^{\alpha \beta} \hat{\delta} g_{\alpha \beta} \mathcal{E}^{\mu \nu} \xi _{\nu} \right)
\end{equation}
where $\hat{\delta}$ denotes variation with respect to the dynamical fields. The off-shell ADT current is defined as \cite{1,2,3}
\begin{equation}\label{8}
  \sqrt{-g} \mathcal{J}^{\mu} _{\text{ADT}} (g , \delta g ; \xi) = \delta \mathcal{E} ^{\mu \nu} \xi_{\nu} + \mathcal{E} ^{\mu \nu} \delta g_{\nu \lambda} \xi ^{\lambda} - \frac{1}{2} \xi ^{\mu} \mathcal{E} ^{\alpha \beta} \delta g_{\alpha \beta} + \frac{1}{2} g ^{\alpha \beta} \delta g_{\alpha \beta} \mathcal{E}^{\mu \nu} \xi _{\nu}.
\end{equation}
The off-shell ADT current $\mathcal{J} ^{\mu} _{\text{ADT}} (g ; \delta g)$ is conserved off-shell for arbitrary field-independent Killing vector field which is admitted by the spacetime everywhere. Also, the symplectic current define as an antisymmetric bilinear map on perturbations \cite{8}
\begin{equation}\label{9}
  \omega ^{\mu} (g; \delta _{1} g , \delta _{2} g) = \delta _{1} \Theta ^{\mu}(g ; \delta _{2} g ) - \delta _{2} \Theta ^{\mu}(g ; \delta _{1} g ) - \Theta ^{\mu}(g ; [\delta _{1},\delta _{2}] g ).
\end{equation}
The above expression for the symplectic current reduces to the Lee-Wald one \cite{4,5,6,7}, namely $\omega ^{\mu}_{\text{LW}} = \delta _{1} \Theta ^{\mu}(g ; \delta _{2} g ) - \delta _{2} \Theta ^{\mu}(g ; \delta _{1} g ) $, when two variations $\delta_{1}$ and $\delta_{2}$ are commute, i.e. $[\delta _{1},\delta _{2}] g=0$. The symplectic current \eqref{9} is conserved on-shell and it gives us conserved charges correspond to asymptotically field-independent Killing vectors. It should be noted that for the case in which $\xi$ is field-dependent we have $ [\hat{\delta},\delta _{\xi}] = \delta _{\hat{\delta} \xi} $, then Eq.\eqref{9} becomes
\begin{equation}\label{10}
  \omega ^{\mu} (g; \hat{\delta} g , \delta _{\xi} g) = \hat{\delta} \Theta ^{\mu}(g ; \delta _{\xi} g ) - \delta _{\xi} \Theta ^{\mu}(g ; \hat{\delta} g ) - \Theta ^{\mu}(g ; \delta _{\hat{\delta} \xi} g ).
\end{equation}
It is easy to see that Eq.\eqref{10} will be reduce to the Lee-Wald symplectic current when $\xi$ is field-independent, i.e. $ \hat{\delta} \xi =0$. In the paper \cite{9}, the authors have generalized the off-shell ADT current as following,
\begin{equation}\label{11}
   \mathcal{J}^{\mu} _{\text{GADT}} (g , \delta g ; \xi) =  \mathcal{J}^{\mu} _{\text{ADT}} (g , \delta g ; \xi) + \frac{1}{2 \sqrt{-g}} \omega ^{\mu} _{\text{LW}} (g; \delta g , \delta _{\xi} g),
\end{equation}
this current is conserved off-shell for the asymptotically field-independent Killing vector fields as well as field-independent Killing vector fields admitted by spacetime everywhere.\\
For the case in which $\xi$ depends on the dynamical fields, it seems to be sensible replacing $\delta$ and the Lee-Wald symplectic current by $\hat{\delta}$ and $\omega ^{\mu} (g; \hat{\delta} g , \delta _{\xi} g)$ in Eq.\eqref{11}, respectively. Thus, we can define the extended off-shell ADT current as
\begin{equation}\label{12}
   \mathfrak{J}^{\mu} _{\text{ADT}} (g , \hat{\delta} g ; \xi) =  \mathcal{J}^{\mu} _{\text{ADT}} (g , \hat{\delta} g ; \xi) + \frac{1}{2 \sqrt{-g}} \omega ^{\mu} (g; \hat{\delta} g , \delta _{\xi} g).
\end{equation}
It is clear that the extended off-shell ADT current $\mathfrak{J}^{\mu} _{\text{ADT}}$ is conserved off-shell for asymptotically field dependent Killing vector fields as well as field-dependent Killing vector fields admitted by spacetime everywhere. By considering Eq.\eqref{12}, one can rewrite Eq.\eqref{7} as follows:
\begin{equation}\label{13}
   \sqrt{-g} \mathfrak{J}^{\mu} _{\text{ADT}} (g , \hat{\delta} g ; \xi) = \partial _{\nu} \left[ \sqrt{-g} \mathcal{Q} _{\text{ADT}} ^{\nu \mu} (g , \hat{\delta} g ; \xi) \right],
\end{equation}
where $\mathcal{Q} _{\text{ADT}} ^{\mu \nu} (g , \hat{\delta} g ; \xi)$ is defined as the extended off-shell ADT charge and it is given by
\begin{equation}\label{14}
  \sqrt{-g} \mathcal{Q}_{\text{ADT}} ^{\mu \nu} (g , \hat{\delta} g ; \xi) = \frac{1}{2}\hat{\delta} K^{\mu \nu}(g ; \xi) - \frac{1}{2} K^{\mu \nu}(g ; \hat{\delta} \xi)- \xi ^{[\mu} \Theta ^{\nu]}(g ; \hat{\delta} g ).
\end{equation}
By substituting $K^{\mu \nu}= \sqrt{-g} \tilde{K}^{\mu \nu}$ and $\Theta ^{\mu}= \sqrt{-g} \tilde{\Theta }^{\mu}$ into Eq.\eqref{14} we have
\begin{equation}\label{15}
\begin{split}
   \mathcal{Q}_{\text{ADT}} ^{\mu \nu} (g , \hat{\delta} g ; \xi) = & \frac{1}{2}\hat{\delta} \tilde{K}^{\mu \nu}(g ; \xi) + \frac{1}{4} g^{\alpha \beta} \hat{\delta} g_{\alpha \beta} \tilde{K}^{\mu \nu}(g ; \xi) \\
     & - \frac{1}{2} \tilde{K}^{\mu \nu}(g ; \hat{\delta} \xi)- \xi ^{[\mu} \tilde{\Theta} ^{\nu]}(g ; \hat{\delta} g ).
\end{split}
\end{equation}
The above expression for the extended off-shell ADT charge reduce to the generalized ADT charge \cite{9} when $\xi$ is field-independent, i.e. $\hat{\delta} \xi =0$. Now, we can define the perturbation of the conserved charge by integrating from the extended off-shell ADT charge over a spacelike codimension two surface
\begin{equation}\label{16}
  \hat{\delta} Q(\xi) = \frac{1}{16 \pi G} \int _{\Sigma} (d^{D-2} x) _{\mu \nu} \sqrt{-g} \mathcal{Q}_{\text{ADT}} ^{\mu \nu} (g , \hat{\delta} g ; \xi),
\end{equation}
where
\begin{equation}\label{17}
  (d^{D-2} x) _{\mu \nu} = \frac{1}{2(D-2)!} \varepsilon _{\mu \nu \alpha _{1} \cdots \alpha _{D-2}} dx^{\alpha_{1}} \cdots dx^{\alpha_{D-2}}
\end{equation}
The charge defined by Eq.\eqref{16} is conserved off-shell for the asymptotically field-dependent Killing vector fields as well as field-dependent Killing vector fields admitted by spacetime everywhere.\\
As we mentioned earlier, the Lagrangian of a generally covariant theory of gravity is given by $\mathcal{L} = \mathcal{L} (g_{\mu \nu} , R, R_{\mu \nu} R^{\mu \nu},\dots)$, so, we have the following expressions for $ \tilde{K}^{\mu \nu} $ and $ \tilde{\Theta} ^{\mu}$ \cite{3}
\begin{equation}\label{18}
\begin{split}
 \tilde{K}^{\mu \nu} (g ; \xi)  = & 2 P^{\mu \nu \alpha \beta} \nabla _{\alpha} \xi _{\beta} - 4 \xi _{\beta} \nabla _{\alpha} P^{\mu \nu \alpha \beta},\\
  \tilde{\Theta} ^{\mu} (g ; \hat{\delta} g) = & 2 \left( P^{\mu \alpha \beta \nu} \nabla _{\nu} \hat{\delta} g_{\alpha \beta} - \hat{\delta} g_{\alpha \beta} \nabla _{\nu} P^{\mu \alpha \beta \nu} \right),
  \end{split}
\end{equation}
where $P^{\mu \nu \alpha \beta}= \partial \mathcal{L} / \partial R _{\mu \nu \alpha \beta}$. For the Einstein gravity, we have
\begin{equation}\label{19}
  P^{\alpha \mu \beta \nu} = \frac{1}{2} \left( g^{\mu \nu} g^{\alpha \beta} - g^{\alpha \nu} g^{\mu \beta} \right),
\end{equation}
therefore in this case Eq.\eqref{18} will be reduced to
\begin{equation}\label{20}
\begin{split}
  \tilde{K}^{\mu \nu} (g ; \xi) =& 2 \nabla ^{[\mu} \xi ^{\nu]},\\
  \tilde{\Theta} ^{\mu} (g ; \hat{\delta} g) =& \nabla ^{\alpha} \left( g^{\mu \beta} \hat{\delta} g_{\alpha \beta} \right) - \nabla ^{\mu} \left( g^{\alpha \beta} \hat{\delta} g_{\alpha \beta} \right).
  \end{split}
\end{equation}
By substituting Eq.\eqref{20} into Eq.\eqref{15} we find that \footnote{It is easy to see that the Abbott-Deser formula (see, equation (2.1) of \cite{6'}) has two more terms than the formula \eqref{21}.}
\begin{equation}\label{21}
\begin{split}
  \mathcal{Q}_{\text{ADT}} ^{\mu \nu} (g , \hat{\delta} g ; \xi) = & - h^{\lambda [ \mu} \nabla _{\lambda} \xi ^{\nu]} + \xi ^{\lambda} \nabla ^{[\mu} h^{\nu]}_{\lambda} + \frac{1}{2} h \nabla ^{[\mu} \xi ^{\nu]}\\
  & - \xi ^{[\mu} \nabla _{\lambda} h^{\nu] \lambda} + \xi ^{[\mu} \nabla^{\nu]}h,
\end{split}
\end{equation}
where $h_{\mu \nu}= \hat{\delta} g_{\mu \nu}$. Although this formula is independent of $\hat{\delta} \xi$, we have shown that this formula is valid for the case in which $\xi$ is field-dependent.
\subsection{An example}
Let $x^{\mu}=(u,r,\theta,\phi)$ and $A,B,\dots=2,3$. We consider an asymptotically flat spacetime presented in \cite{11}. The line-element of an asymptotically flat spacetime can be written in the following form \cite{10,11}
\begin{equation}\label{22}
  ds^{2}= e^{2X} \frac{V}{r} du^{2} - 2 e^{2X} du dr + g_{AB} (dx^{A} - U^{A} du) (dx^{B} - U^{B} du),
\end{equation}
where $X$, $V$ and $g_{AB} (\text{det} g_{AB})^{-1/2} $ are $6$ functions of the coordinates. Also, we assume that $g^{AC} g_{CB}= \delta ^{A}_{B}$ and we impose following gauge conditions
\begin{equation}\label{23}
  g_{rr}=0, \hspace{1 cm} g_{rA}=0,
\end{equation}
\begin{equation}\label{24}
  \partial _{r} \left( r^{-4} \text{det} g_{AB} \right) = 0.
\end{equation}
The line-element \eqref{22} solves the Einstein field equations when $g_{AB}$, $X$, $V$ and $U^{A}$ are given as
\begin{equation}\label{25}
  g_{AB}=r^{2} \bar{\gamma}_{AB}+ r \bar{C}_{AB} + \bar{D}_{AB} + \frac{1}{4} \bar{\gamma}_{AB} \bar{C}^{C}_{D} \bar{C}^{D}_{C} + \mathcal{O}(r^{-\varepsilon}),
\end{equation}
\begin{equation}\label{26}
  X=-\frac{1}{32r^{2}} \bar{C}^{A}_{B} \bar{C}^{B}_{A}- \frac{1}{12r^{3}} \bar{C}^{A}_{B} \bar{D}^{B}_{A}+\mathcal{O}(r^{-3-\varepsilon})
\end{equation}
\begin{equation}\label{27}
  \frac{V}{r} = - \frac{1}{2} \bar{R} + \frac{2 M}{r} + \mathcal{O}(r^{-1-\varepsilon})
\end{equation}
\begin{equation}\label{28}
\begin{split}
   U^{A}= & -\frac{1}{2r^{2}} \bar{\nabla}_{B} \bar{C}^{AB} \\
     & -\frac{2}{3r^{3}} \left[ \left( \ln r + \frac{1}{3} \right) \bar{\nabla}_{B} \bar{D}^{AB} - \frac{1}{2} \bar{C}^{A}_{B} \bar{\nabla}_{C} \bar{C}^{CB} + \bar{N}^{A} \right] + \mathcal{O}(r^{-2-\varepsilon})
\end{split}
\end{equation}
where $\bar{\gamma}^{AC} \bar{\gamma}_{CB}= \delta ^{A}_{B}$, $\bar{\gamma}_{AB} dx^{A} dx^{B} = e^{2 \varphi(x^C) } (d \theta ^{2} + \sin ^{2} \theta d \phi ^{2})$ and indices on $\bar{C}_{AB}$ and $\bar{D}_{AB} $ are raised with $\bar{\gamma}^{AB}$. Also, $ \bar{C}^{C}_{C}=\bar{D}^{C}_{C} = \partial _{u} \bar{D}_{AB} =0$. $\bar{\nabla}_{A}$ is the covariant derivative associated to $\bar{\gamma}_{AB}$ and $\bar{R}$ is the scalar curvature of $\bar{\nabla}_{A}$. In Eq.\eqref{27}, $M=M(u,x^{A})$ is the mass aspect and, in Eq.\eqref{28}, $\bar{N}^{A}=\bar{N}^{A}(u,x^{B})$ is the angular momentum aspect. We should mention that $\bar{C}_{AB}$, $\bar{D}_{AB}$, $M$ and $\bar{N}^{A}$ are dynamical fields. The metric under transformations generated by $\xi$ transforms as $\delta_{\xi} g_{\mu \nu} = \pounds_{\xi} g_{\mu \nu}$. The variation generated by the  following vector field preserves the fall-off conditions \eqref{25}-\eqref{28}
\begin{equation}\label{29}
  \begin{split}
     \xi^{u} = & f \\
       \xi ^{r} = & -\frac{r}{2} (\bar{\nabla}_{A} \xi^{A} - f_{,B} U^{B})\\
       \xi ^{A} = & Y^{A} +I^{A} \hspace{0.5 cm} ; \hspace{0.5 cm} I^{A}= - f_{,B} \int_{r}^{\infty} dr^{\prime} (e^{2X} g^{AB})
  \end{split}
\end{equation}
where $f=e^{\varphi} T + \frac{1}{2} u \psi$, $Y^{A}=Y^{A}(x^{B})$ , $T=T(x^{B})$ and $\psi = \bar{\nabla}_{A} Y^{A}$, where $Y^{A}$ is a conformal Killing vector of $\bar{\gamma}_{AB}$. Preserving the boundary conditions means that the metric $g_{\mu \nu}(\Phi)$ is mapped into $g_{\mu \nu}(\Phi+\delta_{\xi} \Phi )$ by transforms generated by $\xi$, where $\Phi$ is the collection of dynamical fields. The change of dynamical fields $ \delta_{\xi} \Phi $ under transformation generated by $\xi$ are given in the paper \cite{6'}.\\
Now, we take codimension two surface $\Sigma$ to be a $(u,r)$-constant surface. Thus, $ru$-component of the extended ADT charge is important. We can rewrite $ru$-component of the extended ADT charge \eqref{21} as
\begin{equation}\label{30}
\begin{split}
   2 \mathcal{Q}_{\text{ADT}} ^{r u} = & \xi ^{r} \left[ \nabla ^{r} h^{u}_{r} + \nabla^{u} h - \nabla ^{u} h^{r}_{r} - \nabla _{\lambda} h^{u\lambda} \right] \\
     & + \xi^{u} \left[ \nabla ^{r} h^{u}_{u} - \nabla^{r} h - \nabla ^{u} h^{r}_{u} + \nabla _{\lambda} h^{r \lambda} \right] \\
     & + \xi^{A} \left[ \nabla ^{r} h^{u}_{A} - \nabla ^{u} h^{r}_{A} \right] + \frac{1}{2} h \left( \nabla ^{r} \xi ^{u} - \nabla ^{u} \xi ^{r} \right) \\
     & + h^{\lambda u} \nabla _{\lambda} \xi ^{r} - h^{\lambda r} \nabla _{\lambda} \xi ^{u}.
\end{split}
\end{equation}
Two last terms in the right hand side of Eq.\eqref{30} can be simplified as
\begin{equation}\label{31}
\begin{split}
   \frac{1}{2} h \left( \nabla ^{r} \xi ^{u} - \nabla ^{u} \xi ^{r} \right)+ h^{\lambda u} \nabla _{\lambda} \xi ^{r} - h^{\lambda r} \nabla _{\lambda} \xi ^{u} = & \left(\frac{1}{2} h g^{ru} - h^{ru} \right) \left( \nabla _{u} \xi ^{u} - \nabla _{r} \xi^{r} \right) \\
     & + \left(\frac{1}{2} h g^{rA} - h^{rA} \right) \nabla _{A} \xi^{u},
\end{split}
\end{equation}
where we have used the equation $\nabla _{r} \xi ^{u} = 0$. Since $h^{\mu \nu}=g^{\mu \alpha} g^{\nu \beta} \hat{\delta} g_{\alpha \beta} = - \hat{\delta} g^{\mu \nu}$, then by considering line-element \eqref{22}, we find that
\begin{equation}\label{32}
 \frac{1}{2} h \left( \nabla ^{r} \xi ^{u} - \nabla ^{u} \xi ^{r} \right)+ h^{\lambda u} \nabla _{\lambda} \xi ^{r} - h^{\lambda r} \nabla _{\lambda} \xi ^{u} = - e^{-2X} \hat{\delta} U^{A} \nabla _{A} \xi^{u}.
\end{equation}
By substituting Eq.\eqref{25} and Eq.\eqref{26} into Eq.\eqref{29}, we have
\begin{equation}\label{33}
  \xi ^{A} = Y^{A} - \frac{1}{r} \bar{\nabla}^{A}f+\frac{1}{2r^{2}} \bar{C}^{AB}\bar{\nabla}_{B} f + \mathcal{O} (r^{-3}),
\end{equation}
and by some calculations one can show that
\begin{equation}\label{34}
  \nabla _{A} \xi^{u} = r Y_{A} + \frac{1}{2} \bar{C}_{AB} Y^{B} + \mathcal{O} (r^{-1}),
\end{equation}
\begin{equation}\label{35}
  \xi^{r} = -\frac{1}{2} r \psi + \frac{1}{2} \bar{\Box} f + \mathcal{O} (r^{-1}).
\end{equation}
By substituting Eq.\eqref{34} and Eq.\eqref{28} into Eq.\eqref{32} we find that
$$r^{2} e^{2X}\left[ \frac{1}{2} h \left( \nabla ^{r} \xi ^{u} - \nabla ^{u} \xi ^{r} \right)+ h^{\lambda u} \nabla _{\lambda} \xi ^{r} - h^{\lambda r} \nabla _{\lambda} \xi ^{u} \right] =\frac{1}{2} r Y_{A} \bar{\nabla} _{B} \hat{\delta} \bar{C}^{AB}$$
$$ + \left[ \frac{2}{3} \left( \ln r + \frac{1}{3} \right) Y_{A} \bar{\nabla} _{B} \hat{\delta} \bar{D}^{AB} - \frac{1}{3} Y_{A}\hat{\delta} \left(\bar{C}^{A}_{B} \bar{\nabla} _{C} \bar{C}^{BC}\right) +\frac{2}{3} Y_{A} \hat{\delta} \bar{N}^{A} \right] $$
\begin{equation}\label{36}
 + \frac{1}{4} \bar{\nabla} _{B} \hat{\delta} \bar{C}^{AB} \bar{C}_{AC}Y^{C} + \mathcal{O} (r^{-1}) .
\end{equation}
Also, we have \cite{6'}
\begin{equation}\label{37}
  \nabla ^{r} h^{u}_{r} + \nabla^{u} h - \nabla ^{u} h^{r}_{r} - \nabla _{\lambda} h^{u\lambda}= \frac{1}{4r^{3}} \bar{C}^{AB} \hat{\delta} \bar{C}_{AB} + \mathcal{O} (r^{-3-\varepsilon}),
\end{equation}
\begin{equation}\label{38}
\begin{split}
      \nabla ^{r} h^{u}_{u} - \nabla^{r} h - \nabla ^{u} h^{r}_{u} + \nabla _{\lambda} h^{r \lambda} =& \frac{1}{r^{2}} \left[4 \hat{\delta} M - \frac{1}{2} \bar{\nabla} _{A} \bar{\nabla} _{B} \hat{\delta} \bar{C}^{AB} + \frac{1}{2} \partial _{u} \bar{C}^{AB} \hat{\delta} \bar{C}_{AB} \right]\\
     & + \mathcal{O} (r^{-2-\varepsilon}),
\end{split}
\end{equation}
\begin{equation}\label{39}
  \begin{split}
     \nabla ^{r} h^{u}_{A} - \nabla ^{u} h^{r}_{A} = & \frac{1}{2r} \bar{\nabla} _{B} \hat{\delta} \bar{C}^{B}_{A} + \frac{2}{3r^{2}} \left( 2 \ln r - \frac{1}{3} \right) \bar{\nabla} _{B} \hat{\delta} \bar{D}^{B}_{A} \\
       & + \frac{1}{r^{2}} \left[ \frac{4}{3} \hat{\delta} \bar{N}_{A} +\frac{1}{3} \hat{\delta} \left(\bar{C}_{AB} \bar{\nabla} _{C} \bar{C}^{BC}\right) - \frac{1}{4} \bar{C}_{AB} \bar{\nabla} _{C} \hat{\delta} \bar{C}^{BC} \right]\\
       & + \mathcal{O} (r^{-2-\varepsilon}).
  \end{split}
\end{equation}
As we mentioned earlier, we take codimension two surface $\Sigma$ to be a $(u,r)$-constant surface so Eq.\eqref{16} becomes
\begin{equation}\label{40}
  \not\hat{\delta} Q(\xi) = \frac{1}{8 \pi G} \int _{2\text{-sphere}} d^{2} \Omega \left( r^{2} e^{2X} \mathcal{Q}_{\text{ADT}} ^{r u} \right) ,
\end{equation}
where
\begin{equation}\label{41}
  d^{2} \Omega = e^{2 \varphi} \sin \theta d \theta d \phi.
\end{equation}
The symbol $\not\hat{\delta}$ emphasizes that the perturbation of conserved charge may be non-integrable. By substituting Eqs.\eqref{36}-\eqref{39} into Eq.\eqref{40} we have
\begin{equation}\label{42}
  \begin{split}
  \not\hat{\delta} Q(\xi) = \frac{1}{16 \pi G} \int _{2\text{-sphere}} d^{2} \Omega  & \{r Y_{A}  \bar{\nabla} _{B} \hat{\delta}\bar{C}^{AB} - \frac{1}{2} \bar{\nabla} _{A} f \bar{\nabla} _{B} \hat{\delta}\bar{C}^{AB} \\
       & - \frac{1}{8} \psi \bar{C}^{AB} \hat{\delta} \bar{C}_{AB} + 2 \ln r Y_{A}\bar{\nabla}_{B} \hat{\delta}\bar{D}^{AB} \\
       & + 2 Y_{A} \hat{\delta}\bar{N}^{A} + 4 f \hat{\delta} M - \frac{1}{2} f \bar{\nabla} _{A} \bar{\nabla} _{B} \hat{\delta}\bar{C}^{AB}\\
       & + \frac{1}{2} f \partial _{u} \bar{C}^{AB} \hat{\delta} \bar{C}_{AB}\}.
  \end{split}
\end{equation}
Using the conformal Killing equation for the $Y^{A}$, i.e. $\bar{\nabla} _{(A} Y_{B)} = \frac{1}{2} \bar{\gamma}_{AB} \bar{\nabla}_{C} Y^{C}$, and integrations by parts, Eq.\eqref{42} can be simplified to
\begin{equation}\label{43}
  \begin{split}
  \not\hat{\delta} Q(\xi) =  & \frac{1}{16 \pi G} \hat{\delta} \int _{2\text{-sphere}} d^{2} \Omega \left\{ 4 f M + Y^{A} \left[ 2 \bar{N}_{A} + \frac{1}{16} \partial _{A} \left( \bar{C}^{BC} \bar{C}_{BC} \right) \right] \right\}\\
  &+ \frac{1}{16 \pi G} \int _{2\text{-sphere}} d^{2} \Omega \left[ \frac{1}{2} f \partial _{u} \bar{C}^{AB} \hat{\delta} \bar{C}_{AB}\right].
  \end{split}
\end{equation}
In the right hand side of Eq.\eqref{43}, the first term is the integrable part of the surface charge and the second term is the non-integrable part. The conserved charge perturbation \eqref{43} which we found in this way is exactly matched with that of the paper \cite{6'}. Here we have obtained this result by a different approach. As has been discussed in \cite{6'} these charges are as representations of the
$BMS_4$ symmetry algebra.
\section{Near horizon conserved charges of non-extremal black holes as Neother charges}
Consider near horizon metric of a non-extremal black hole in the Eddington-Filkenstein coordinates. Let $v$ and $\rho$ are the advanced time and radial coordinate, respectively. We suppose that the event horizon (which is a null surface) located at $\rho=0$ and it is a non-expanding surface. Following \cite{13,14,15}, components of the metric close to Near horizon region behave like
\begin{equation}\label{44}
  g_{\mu \nu} = \left(
    \begin{array}{ccc}
      -2 \kappa \rho + \mathcal{O}(\rho ^{2})& 1 & \rho N_{B} + \mathcal{O}(\rho ^{2}) \\
      1 & 0 & 0 \\
      \rho N_{A} + \mathcal{O}(\rho ^{2}) & 0 & \Omega _{AB} + \rho \lambda _{AB} + \mathcal{O}(\rho ^{2})\\
    \end{array}
  \right),
\end{equation}
where $\kappa$, $N_{A}$, $\Omega _{AB}$ and $\lambda _{AB}$ are functions of $(v,x^{A})$. As before, we assume that $x^{A}=(\theta , \phi)$. The inverse of metric \eqref{45} is given by
\begin{equation}\label{45}
  g^{\mu \nu} = \left(
    \begin{array}{ccc}
      0 & 1 & 0 \\
      1 & 2 \kappa \rho + \mathcal{O}(\rho ^{2}) & - \rho N^{B} + \mathcal{O}(\rho ^{2}) \\
      0 & - \rho N ^{A} + \mathcal{O}(\rho ^{2}) & \Omega ^{AB} - \rho \lambda ^{AB} + \mathcal{O}(\rho ^{2})\\
    \end{array}
  \right),
\end{equation}
where $\Omega ^{AB}$ is the inverse of $\Omega _{AB}$ and, in this section, $A,B,\dots$ indices are raised with $\Omega^{AB}$. By some calculations, one can show that the metric connections correspond to the metric \eqref{44} are
\begin{equation}\label{46}
  \begin{split}
       & \Gamma ^{\lambda}_{\rho \rho} = 0, \hspace{0.4 cm} \Gamma ^{v}_{\lambda \rho} = 0,\hspace{0.4 cm} \Gamma ^{\rho}_{A \rho} =  \frac{1}{2} N_{A} + \mathcal{O}(\rho), \hspace{0.4 cm} \Gamma ^{A}_{B \rho} =  \frac{1}{2} \lambda ^{A}_{B} + \mathcal{O}(\rho), \\
       & \Gamma ^{v}_{A B} =  - \frac{1}{2} \lambda _{AB} + \mathcal{O}(\rho), \hspace{0.4 cm} \Gamma ^{A}_{ B C} =  \tilde{\Gamma} ^{A}_{ B C} + \mathcal{O}(\rho), \hspace{0.4 cm} \Gamma ^{A}_{\rho v}= \frac{1}{2} N^{A} + \mathcal{O}(\rho), \\
       & \Gamma ^{v}_{v A}= -\frac{1}{2} N_{A},\hspace{0.4 cm} \Gamma ^{\rho}_{\rho v}= -\kappa + \mathcal{O}(\rho),\hspace{0.4 cm} \Gamma ^{A}_{B v}= \frac{1}{2} \Omega^{AC} \Omega _{CB,v} + \mathcal{O}(\rho), \\
       & \Gamma ^{v}_{v v}= \kappa, \hspace{0.4 cm} \Gamma ^{\rho}_{AB}= -\frac{1}{2} \Omega _{AB,v} + \mathcal{O}(\rho), \hspace{0.4 cm} \Gamma ^{\rho}_{A v}= \mathcal{O}(\rho), \hspace{0.4 cm} \Gamma ^{\rho}_{v v}= \mathcal{O}(\rho),\\
       & \Gamma ^{A}_{v v}= \mathcal{O}(\rho),
  \end{split}
\end{equation}
where $\tilde{\Gamma} ^{A}_{ B C}$ is the connection associated to $\Omega_{AB}$.\\
The variation generated by the following vector field preserves the fall-off conditions \eqref{44}
\begin{equation}\label{47}
  \begin{split}
      \chi ^{v} = & f, \\
      \chi ^{\rho} = & - \rho \partial _{v} f +\frac{1}{2} \rho ^{2} N^{A} \partial _{A} f + \mathcal{O}(\rho ^{3}) \\
       \chi ^{A} = & Y^{A} + \rho \partial ^{A} f + \frac{1}{2} \rho ^{2} \lambda ^{AB} \partial _{B} f + \mathcal{O}(\rho ^{3}),
  \end{split}
\end{equation}
where $f=f(v,x^{A})$ and $Y^{A} = Y^{A} (x^{B})$ \cite{15}. Here, as we mentioned in the previous section, the metric under transformation generated by $\chi$ transforms as $\delta_{\chi} g_{\mu \nu} = \pounds_{\chi} g_{\mu \nu}$. In this section, we consider the gauge conditions $g_{\rho \rho}=g_{\rho A}=0$ and $g_{\rho v}=1$. Due to these gauge conditions the $v$ component of $\chi$ has been fixed exactly. Also, the fall-off conditions \eqref{44} are considered up to $\mathcal{O}(\rho^{2})$ so to preserve these boundary conditions under transformations generated by $\chi$, we need to consider $\chi ^{\rho}$ and $\chi ^{A}$ up to $\mathcal{O}(\rho ^{3})$ (see ref. \cite{15}).\\
In section 2, we have showed that $J^{\mu}$ is a Noether current density which is conserved off-shell for any vector field $\xi$, see equations \eqref{4} and \eqref{5}. Then, by virtue of Poincare lemma, we have introduced Noether charge density \eqref{6}, namely $K^{\mu \nu} (\xi)= \sqrt{-g} \tilde{K}^{\mu \nu}(\xi)$ where $ \tilde{K}^{\mu \nu}(\xi) $ is given by Eq.\eqref{20}. Now, we can define the near horizon conserved charge by integrating form the Noether charge density over event horizon
\begin{equation}\label{48}
  \mathfrak{Q} (\chi) = \frac{1}{8 \pi G} \lim _{\rho \rightarrow 0} \int_{2-\text{sphere}} (d^{D-2} x) _{\mu \nu} \sqrt{-g} \nabla ^{[\mu} \chi ^{\nu]},
\end{equation}
where we have replaced $\xi$ by $\chi$. By substituting equations \eqref{44}-\eqref{47} into the Eq.\eqref{48}, we find that
\begin{equation}\label{49}
  \mathfrak{Q} (f,Y^{A}) = \frac{1}{16 \pi G} \int_{\text{Horizon}} d^{2}x \sqrt{\text{det} \Omega _{AB}} \left[  2 \partial _{v} f + 2 \kappa f - Y^{A} N_{A} \right],
\end{equation}
In the stationary case, in which metric and $f$ are independent of the advanced time, Eq.\eqref{49} reduced to
\begin{equation}\label{50}
  \mathfrak{Q} (f,Y^{A}) = \frac{1}{16 \pi G} \int_{\text{Horizon}} d^{2}x \sqrt{\text{det} \Omega _{AB}} \left[ 2 \kappa f - Y^{A} N_{A} \right].
\end{equation}
This result is exactly what has been found in the paper \cite{15}, where the authors have considered the additional restrictions demanded by integrability (see Eq.(37) in the paper \cite{15}).
As we have mentioned in the introduction, the near horizon geometry of non-extermal black hole solutions of a generally covariant theory of gravity exhibits an infinite-dimensional symmetry which is not exactly $BMS_4$.  The authors of \cite{15} have shown that the full symmetry algebra  include two sets of supertranslations in semi-direct sum with two mutually commuting copies of Virasoro algebras. So, at the horizon we have found the so-called $BMS^H_4$ algebra provided in reference \cite{15}.
\section{Conclusion}
We have considered a generally covariant theory of gravity, and found an off-shell conserved current Eq.\eqref{5} by virtue of the Bianchi identities. In order to define an extended off-shell ADT conserved current, we took variation of off-shell conserved current Eq.\eqref{5} with respect to the dynamical fields. The generalized off-shell ADT current Eq.\eqref{11} is conserved for the asymptotically field-independent Killing vector fields and field-independent Killing vector fields admitted by spactime everywhere. We have extended the generalized off-shell ADT current by replacing the Lee-Wald symplectic current by the symplectic current Eq.\eqref{10} to define the extended off-shell ADT current Eq.\eqref{12} which is conserved for asymptotically field-dependent Killing vector fields and field-dependent Killing vector fields admitted by spactime everywhere as well as field-independent one. Using the extended off-shell ADT current Eq.\eqref{12} we have defined extended off-shell ADT charge Eq.\eqref{15}. Also, we have defined the conserved charge perturbation Eq.\eqref{16} by integrating of the extended off-shell ADT charge Eq.\eqref{15} over a spacelike codimension two surface. Then we applied the presented formalism to find conserved charge perturbation of an asymptotically flat spacetime which is presented in the paper \cite{11}. The obtained result Eq.\eqref{43} is exactly matched with the result of the paper \cite{6'}. Then we have considered the fall-off conditions Eq.\eqref{44} near to the horizon of a non-extremal black hole. The considered fall-off conditions Eq.\eqref{44} are preserved under variation of metric due to a diffeomorphism generated by the vector field Eq.\eqref{47}. We have found near horizon conserved charges of considered fall-off conditions Eq.\eqref{44} associated to the vector field Eq.\eqref{47} as the Noether charges.
\section{Acknowledgments}
M. R. Setare  thanks Dr. A. Sorouri  for his help in improvement the English of the text.


\begin{thebibliography}{9}
\bibitem{6'}G. Barnich, C. Troessaert, JHEP 12, 105 (2011).
\bibitem{15} L. Donnay, G. Giribet, H. A. González, M. Pino,  arXiv:1607.05703 [hep-th].
\bibitem{1'}H. Bondi, M. G. J. van der Burg, and A. W. K.
Metzner, Royal Society of London Proceedings Series A 269, (Aug., 1962) 21-52.
 \bibitem{2'}R. K. Sachs,
Proceedings of the Royal Society of London Series A 270 (Oct., 1962) 103-126.
\bibitem{10}R. K. Sachs,  Phys. Rev. 128, (1962) 2851.
\bibitem{4'}G. Barnich and G. Compere,  Class. Quant. Grav. 24
(2007) F15. Corrigendum: ibid 24 (2007) 3139.
\bibitem{5'}G. Barnich, C. Troessaert, Proceedings of the Workshop on Non Commutative Field Theory and Gravity, September 8-12, 2010.
\bibitem{7'}S. Hollands and A. Ishibashi, J. Math. Phys. 46, 022503, (2005).
\bibitem{8'}K. Tanabe, N. Tanahashi, and T. Shiromizu, J. Math. Phys. 51, 062502, (2010).
\bibitem{9'}A. Ashtekar, Phys. Rev. Lett. 46 (1981) 573.
\bibitem{10'}A. Ashtekar, "Asymptotic Quantization: Based on 1984 Naples Lectures,". Naples,
Italy: Bibliopolis (1987) 107 p. (Monographs and textbooks in physical science, 2).
\bibitem{11'}L. Donnay, G. Giribet, H. A. Gonzalez, M. Pino,  Phys. Rev. Lett. 116, 091101 (2016).
\bibitem{b}R. G. Cai, S. M. Ruan, Y. L. Zhang, JHEP 09 (2016) 163.
\bibitem{30'}M. R. Setare, H. Adami, Phys. Lett. B760, 411, (2016).
\bibitem{200'} M. R. Setare, Nucl. Phys. B 898, 259 (2015).
\bibitem{12'}A. Strominger, JHEP 07 (2014) 152.
\bibitem{13'}L. Susskind, AIP Conf. Proc. 493, 98 (1999).
\bibitem{14'}J. Polchinski,  hep-th/9901076.
\bibitem{15'}J. de Boer and S. N. Solodukhin, Nucl. Phys. B665 (2003) 545.
\bibitem{16'}S. N. Solodukhin,  hep-th/0405252.
\bibitem{17'}M. Gary, S. B. Giddings, Phys. Rev D.80, 046008, (2009).
\bibitem{18'}G. Arcioni, C. Dappiaggi, Class. Quant. Grav. 21 (2004) 5655.
\bibitem{19'}G. Arcioni, C. Dappiaggi, Nucl. Phys. B674 (2003) 553.
\bibitem{20'}G. Barnich, C. Troessaert, Phys. Rev. Lett. 105 (2010) 111103.
\bibitem{11} G. Barnich, C. Troessaert, JHEP 05, 062, (2010).
\bibitem{21'}G. Barnich , C. Troessaert, JHEP 1311 (2013) 003; G. Barnich, A. Gomberoff,  H. A. Gonzalez, Phys. Rev. D86 (2012) 024020; G. Barnich, A. Gomberoff  H. A. Gonzalez, Phys. Rev. D87 (2013) 124032; G. Barnich, A. Gomberoff, H. A. Gonzalez, JHEP 05 (2013) 016; G. Barnich, C. Troessaert, JHEP, 03 (2016) 167; B. Oblak, arXiv:1610.08526 [hep-th]; G. Barnich, L. Donnay, J. Matulich, R. Troncoso, 	arXiv:1510.08824 [hep-th]; G. Barnich, B. Oblak,  JHEP 1406 (2014) 129; G. Barnich, B. Oblak,  JHEP 1503 (2015) 033.
 \bibitem{22'}G. Barnich and F. Brandt,  Nucl. Phys. B633, 3 (2002).
\bibitem{23'}G. Barnich, Class. Quant. Grav. 20 (2003) 3685.
\bibitem{24'}G. Barnich and G. Compere, J. Math. Phys. 49 (2008) 042901.
\bibitem{a'} I. Fujisawa, R. Nakayama, Phys. Rev. D \textbf{91} (2015) 126005.
\bibitem{25'}E. E. Flanagan, D. A. Nichols,	arXiv:1510.03386 [hep-th].
\bibitem{1000}L. F. Abbott and S. Deser, Nucl. Phys. B 195, 76 (1982).
\bibitem{1001}S. Deser and B. Tekin, Phys. Rev. Lett. 89, 101101 (2002).
\bibitem{1002}S. Deser and B. Tekin, Phys. Rev. D 67, 084009 (2003).
\bibitem{1} A. Bouchareb and G. Clement, Class. Quant. Grav. \textbf{24} (2007) 5581.
\bibitem{2} S. Nam, J. -D. Park and S. -H. Yi, Phys. Rev. D \textbf{82} (2010) 124049.
\bibitem{3} W. Kim, S. Kulkarni, S.H. Yi, Phys. Rev. Lett. \textbf{111} (2013) 081101.
\bibitem{29} M. R. Setare, H. Adami,  Eur. Phys. J. C 76:187 (2016).
\bibitem{150} M. R. Setare, H. Adami, Nucl. Phys. B \textbf{902} (2016) 115.
\bibitem{40} M. R. Setare, H. Adami, Nucl. Phys. B \textbf{909}, 345 (2016).
\bibitem{8} K. Prabhu, arXiv:1511.00388v2 [gr-qc].
\bibitem{4} R. M. Wald and A. Zoupas, Phys. Rev. D \textbf{61} (2000) 084027.
\bibitem{5} J. Lee and R. M. Wald, J. Math. Phys. \textbf{31} (1990) 725.
\bibitem{6} R. M. Wald, Phys. Rev. D \textbf{48} (1993) 3427.
\bibitem{7} V. Iyer and R. M. Wald, Phys. Rev. D \textbf{50} (1994) 846.
\bibitem{9} S. Hyun, S. A. Park, S. H. Yi, JHEP \textbf{1406} (2014) 151.
\bibitem{13} L. A. Tamburino and J. H. Winicour, Phys. Rev. \textbf{150} (1966) 1039.
\bibitem{14} I. Booth, Phys. Rev. D \textbf{87} (2013) 024008.
\end{thebibliography}
\end{document}